\documentstyle[12pt]{article}
\def\be{\begin{equation}}
\def\ee{\end{equation}}
\def\bea{\begin{eqnarray}}
\def\eea{\end{eqnarray}}

\begin{document}

\begin{center}
{\Large{\bf Closed Superstring in Noncommutative Compact Spacetime}}
\vskip .5cm
{\large Davoud Kamani}
\vskip .1cm
 {\it Institute for Studies in Theoretical Physics and 
Mathematics (IPM)
\\ P.O.Box: 19395-5531, Tehran, Iran}\\
{\it e-mail: kamani@theory.ipm.ac.ir}
\\
\end{center}

\begin{abstract} 
In this paper we study the effects of noncommutativity on a closed 
superstring propagating in the spacetime 
that is compactified on tori. The effects of compactification and 
noncommutativity appear in the momentum, quantization, supercurrent,
super-conformal generators and in the boundary state of 
the closed superstring emitted from a D$_p$-brane with the 
NS$\otimes$NS background $B$-field. 

\end{abstract} 


\newpage
\section{Introduction}

There have been much activities exploring the relation between string theory
and noncommutative geometry \cite{1,2,3}. There have been attempts to explain
noncommutativity on D-brane through the study of open strings
in the presence of background fields \cite{3}.

D-brane can be alternatively described by the ``boundary state'' of the
closed string channel description \cite{4,5,6,7,8,9,10}. 
The boundary state can be interpreted as a source for a closed
string emitted by a D-brane. 
By introducing the background field $B_{\mu \nu}$ in the string
$\sigma$-model one obtains mixed boundary conditions for strings. 
Mixed boundary conditions have been used for studying many properties of
D-branes in the background fields \cite{5,8,9,10}. 

We proceed to study closed superstrings in the spacetime
which is compactified on tori. In a special coordinate system, which we 
call it ``closed string frame'', compactification 
reveals the noncommutativity of
the spacetime through closed strings. In other words, by using the boundary 
state for a closed superstring, we observe the noncommutativity effects on a
D$_p$-brane with background field wrapped on tori.
We shall see that, $O(10,10;{\bf R})$ duality group relates coordinates of
the spacetime and of the closed string frame.

In the first part of this paper, we study the closed 
superstring, its T-duality, its quantization and the noncommutativity 
effects on it in the noncommutative compact spacetime.
In the second part, the boundary state equations of closed superstring 
will be studied.
%%%%%%%%%%%%%%%%%%%%%%%%%%%%%%%%%%%%%%%%%%%%%%%%%%%%%%%%%%%%%%%%%%%%%%%%%%%%
\section{T-duality and closed string frame}

The solution of the equation of motion extracted from the string action 
can be written as the following mode expansion
\bea
X^{\mu}(\sigma , \tau)=X_L^{\mu}(\tau+\sigma)+X_R^{\mu}(\tau-\sigma)\;,
\eea
\bea
&~& X_L^{\mu}(\tau+\sigma) = \xi^{\mu}_L+2\alpha' \Pi^{\mu}_L(\tau+\sigma)
+\frac{i}{2}\sqrt{2\alpha'}\sum_{n\neq 0}\frac{1}{n}
{\tilde a}^{\mu}_n e^{-2in(\tau+\sigma)} \;,
\nonumber\\
&~& X_R^{\mu}(\tau-\sigma) = \xi^{\mu}_R+2\alpha' \Pi^{\mu}_R(\tau-\sigma)
+\frac{i}{2}\sqrt{2\alpha'}\sum_{n\neq 0}\frac{1}{n}
a^{\mu}_n e^{-2in(\tau-\sigma)} \;.
\eea
In the non-zero constant background $B$-field we can write 
\bea
&~& \xi^\mu_L = (1+2\pi \alpha'B)^\mu_{\;\;\nu} x^\nu_L\;,
\nonumber\\
&~& \xi^\mu_R = (1-2\pi \alpha'B)^\mu_{\;\;\nu} x^\nu_R\;,
\eea
\bea
&~& \Pi^\mu_L = (1+2\pi \alpha'B)^\mu_{\;\;\nu} p^\nu_L\;,
\nonumber\\
&~& \Pi^\mu_R = (1-2\pi \alpha'B)^\mu_{\;\;\nu} p^\nu_R\;,
\eea
\bea
&~&{\tilde a}^\mu_n=(1+2\pi \alpha'B)^\mu_{\;\;\nu} {\tilde \alpha}_n^\nu \;,
\nonumber\\
&~& a^\mu_n = (1-2\pi \alpha'B)^{\mu}_{\;\;\nu} \alpha^{\nu}_{n} \;,
\eea
where $B^\mu_{\;\;\;\nu} = g^{\mu\lambda}B_{\lambda\nu}$ and the metric
$g_{\mu\nu}$ also is constant. 
Again the equations (1) and (2) satisfy the equation 
of motion. Assume that all directions of the spacetime are compactified 
on tori. According to the equations (3)-(5) we can write
\bea
&~& X^\mu = {\bar X^\mu} + 2\pi \alpha' B^\mu_{\;\;\nu}{\bar X}'^\nu \;,
\nonumber\\
&~& X'^\mu = {\bar X'^\mu} + 2\pi \alpha' B^\mu_{\;\;\nu}{\bar X}^\nu \;,
\eea
where $X'^\mu$ , the T-dual coordinate of $X^\mu$, has definition
\bea
T_\mu:\; X^\mu(\sigma , \tau) \rightarrow X'^\mu(\sigma , \tau) 
= X^\mu_L- X^\mu_R\;,
\eea
and ${\bar X}^\mu$ and its T-dual coordinate ${\bar X}'^\mu $ are
\bea
{\bar X}^{\mu}(\sigma , \tau)={\bar X}_L^{\mu}(\tau+\sigma)+
{\bar X}_R^{\mu}(\tau-\sigma)\;,
\nonumber\\
{\bar X}'^{\mu} (\sigma , \tau)={\bar X}_L^{\mu}(\tau+\sigma)-
{\bar X}_R^{\mu}(\tau-\sigma)\;.
\eea
Therefore, the mode expansions of the left and the right moving parts of 
${\bar X^\mu}$ and ${\bar X'^\mu}$ are
\bea
&~&{\bar X}_L^{\mu}(\tau+\sigma) = x^{\mu}_L+2\alpha' p^{\mu}_L(\tau+\sigma)
+\frac{i}{2}\sqrt{2\alpha'}\sum_{n\neq 0}\frac{1}{n}
{\tilde \alpha}^{\mu}_n e^{-2in(\tau+\sigma)}\;,
\nonumber\\
&~&{\bar X}_R^{\mu}(\tau-\sigma) = x^{\mu}_R+2\alpha' p^{\mu}_R(\tau-\sigma)
+\frac{i}{2}\sqrt{2\alpha'}\sum_{n\neq 0}\frac{1}{n}
\alpha^{\mu}_n e^{-2in(\tau-\sigma)}\;.
\eea
Let us denote the coordinate system $\{ {\bar X}^\mu \}$ as ``closed string 
frame''.
These coordinates and their T-dual coordinates can be written as
\bea
{\bar X}^\mu = G^{\mu \nu} X_\nu + \frac{1}{2\pi \alpha'} \theta^{\mu \nu}
X'_\nu\;,
\eea
\bea
{\bar X}'^\mu = G^{\mu \nu} X'_\nu + \frac{1}{2\pi \alpha'} \theta^{\mu \nu}
X_\nu\;,
\eea
where there are $X_\mu = g_{\mu\nu}X^\nu$ and $X'_\mu = g_{\mu\nu}X'^\nu$.
Therefore, the coordinates of the closed string frame and their T-dual are
linear combinations of $\{X^\mu \}$ and their T-dual coordinates
$\{X'^\mu \}$. The coefficients of combinations are the elements of the 
open string metric and the noncommutativity parameter
\bea
&~&G^{\mu \nu} = \bigg{(} ( g + 2 \pi \alpha'B)^{-1} g 
( g - 2 \pi \alpha'B )^{-1}\bigg{)}^{\mu \nu}\;, 
\nonumber\\
&~&\theta^{\mu \nu} = -(2\pi \alpha')^2 \bigg{(} ( g 
+ 2 \pi \alpha'B)^{-1} B ( g - 2 \pi \alpha'B )^{-1}\bigg{)}^{\mu \nu}\;. 
\eea
For the ordinary spacetime i.e., $\theta^{\mu \nu}=0$, there are
${\bar X}^\mu = X^\mu$ and ${\bar X}'^\mu = X'^\mu$. Also for the strong
background $B$-field, ${\bar X}^\mu$ is a linear combination of the T-dual
coordinates $\{X'^\mu \}$. In this case ${\bar X}'^\mu$ appears as a
linear combination of $\{X^\mu \}$. In other words, we have
\bea
{\bar X}^\mu = \frac{1}{2\pi \alpha'} (B^{-1})^{\mu \nu} X'_\nu\;,
\nonumber\\
{\bar X}'^\mu = \frac{1}{2\pi \alpha'} (B^{-1})^{\mu \nu} X_\nu\;.
\eea
In the zero slope limit \cite{1} i.e., when $\alpha'$ and the closed string
metric $g_{\mu\nu}$ go to zero like $\alpha' \sim \epsilon^{1/2}$,
$g_{\mu\nu} \sim \epsilon$, where $\epsilon \rightarrow 0$, and $B_{\mu\nu}$
is fixed, we also obtain the above equations.

From the equations (10) and (11) we obtain
\bea
&~& x^\mu = G^{\mu \nu} \xi_\nu + \frac{1}{2\pi \alpha'} \theta^{\mu \nu}
\xi'_\nu\;,
\nonumber\\
&~& x'^\mu = G^{\mu \nu} \xi'_\nu + \frac{1}{2\pi \alpha'} \theta^{\mu \nu}
\xi_\nu\;,
\eea
\bea
&~& \alpha^\mu_n =(G^{-1} - \frac{1}{2\pi \alpha'} \theta)^{\mu \nu}a_{n\nu} \;,
\nonumber\\
&~& {\tilde \alpha}^\mu_n =(G^{-1} + \frac{1}{2\pi \alpha'} \theta)^{\mu \nu}
{\tilde a}_{n\nu} \;,
\eea
where there are $x^\mu = x^\mu_L+x^\mu_R$ , $x'^\mu = x^\mu_L-x^\mu_R$ ,
$\xi^\mu = \xi^\mu_L+\xi^\mu_R$ and $\xi'^\mu = \xi^\mu_L-\xi^\mu_R$.

Imposing the worldsheet supersymmetry in the equation (10) or (11) gives 
the following relations for the worldsheet fermions
\bea
&~& {\bar \psi}^\mu = G^{\mu \nu}\psi_\nu +\frac{1}{2\pi \alpha'} 
\theta^{\mu \nu} \psi'_\nu\;,
\nonumber\\
&~& {\bar \psi}'^\mu = G^{\mu \nu}\psi'_\nu +\frac{1}{2\pi \alpha'} 
\theta^{\mu \nu} \psi_\nu\;,
\eea
where $\psi^\mu = \left( \begin{array}{c} 
\psi^\mu_- \\
\psi^\mu_+
\end{array} \right)$ is worldsheet spinor and
$\psi'^\mu = \left( \begin{array}{c} 
-\psi^\mu_- \\
\psi^\mu_+
\end{array} \right)$ 
is its T-dual spinor. Similar
definitions also hold for the spinors ${\bar \psi}^\mu$ and 
${\bar \psi}'^\mu$. For the ordinary spacetime i.e., $\theta^{\mu \nu} = 0$,  
we have ${\bar \psi}^\mu = \psi^\mu$ and ${\bar \psi}'^\mu = 
\psi'^\mu$. For the strong background $B$-field and also in the zero slope
limit, the spinors $\{{\bar \psi}^\mu \}$ are equivalent to the spinors
$\{\psi'^\mu \}$ and also $\{{\bar \psi}'^\mu \}$ 
are equivalent to $\{\psi^\mu\}$.
Note that $\psi^\mu$ and ${\bar \psi}^\mu$ satisfy 
the equation of motion i.e., 
$\partial_+ \psi^\mu_- =\partial_- \psi^\mu_+ = \partial_+ {\bar \psi}^\mu_- 
=\partial_- {\bar \psi}^\mu_+ = 0$. Furthermore, we have
$\psi_{\pm \mu} = g_{\mu\nu}\psi^\nu_\pm$.

Assume that all coordinates $\{X^\mu\}$ are compacted on tori with radii 
$\{R_{\mu}\}$ therefore,
\bea
X^\mu (\sigma+ \pi , \tau) - X^\mu (\sigma , \tau) = 
2 \pi \Lambda^\mu\;, 
\eea
\bea
\Lambda^\mu = \alpha' ( \Pi^\mu_L - \Pi^\mu_R ) = 
n^\mu R_\mu \;,\;\;\;\;({\rm no\;sum\;on}\; \mu)\,.
\eea
In this case the momentum of the closed string is quantized i.e.,
\bea
\Pi^\mu=\Pi^\mu_L + \Pi^\mu_R = \frac{m^\mu}{R_\mu}\;.
\eea
The integers $n^\mu$ and $m^\mu$ are winding number and momentum
number of the closed string around the 
compact direction $X^\mu$. Therefore, we have the identification
\bea
\xi^\mu \equiv \xi^\mu + 2\pi \Lambda^\mu\;.
\eea

Note that the dual coordinate $X'^\mu$ also is compact 
\bea
X'^\mu (\sigma+ \pi , \tau) - X'^\mu (\sigma , \tau) = 
2 \pi \alpha' \Pi^\mu= 2\pi m^\mu \frac{\alpha'}{R_\mu}\;, 
\eea
therefore, its compactification is on a circle with
radius $\alpha'/R_\mu$ . This compactification gives the identification
\bea
\xi'^\mu \equiv \xi'^\mu + 2\pi \alpha' \Pi^\mu\;.
\eea
From the identifications (20) and (22) we obtain the following
identifications
\bea
&~&x^\mu \equiv x^\mu + 2\pi L^\mu\;,
\nonumber\\
&~& x'^\mu \equiv x'^\mu + 2\pi \alpha' p^\mu\;.
\eea
These imply that the coordinates ${\bar X}^\mu$ and ${\bar X}'^\mu$ also are
compact. The equation (10) or (11) gives $L^\mu$ and $p^\mu$ as
\bea
&~&L^\mu =G^{\mu \nu} \Lambda_\nu + \frac{1}{2\pi} \theta^{\mu \nu}\Pi_\nu\;,
\nonumber\\
&~& p^\mu = G^{\mu \nu} \Pi_\nu + \frac{1}{2\pi\alpha'^2} \theta^{\mu \nu}
\Lambda_\nu\;.
\eea
Therefore, $L^\mu$ and $p^\mu$ are linear combinations of the closed string
momentum numbers and winding numbers. The fact that $L^\mu$ depends on the 
momentum numbers and $p^\mu$ depends on the winding numbers, are consequences 
of the noncommutativity.
Under the T-duality there is the exchange $\Pi^\mu \leftrightarrow 
\frac{1}{\alpha'} \Lambda^\mu\;,$ which leads to the exchange
\bea
T_\mu:\; p^\mu \leftrightarrow \frac{1}{\alpha'} L^\mu\;. 
\eea

The light-cone components of the worldsheet supercurrent have the forms
\bea
&~& J_+ = G_{\mu \nu}{\bar \psi}^\mu_+ \partial_+ {\bar X}^\nu\;,
\nonumber\\
&~& J_- = G_{\mu \nu}{\bar \psi}^\mu_- \partial_- {\bar X}^\nu\;.
\eea
Also the superconformal generators for the R$\otimes$R sector of superstring
are
\bea
&~& L^{(\alpha , d)}_m = \frac{1}{2} \sum_{n\in Z}G_{\mu \nu}:
\alpha^\mu_{m-n}
\alpha^\nu_n : +\frac{1}{4} \sum_{n\in Z}(2n-m)G_{\mu \nu}:d^\mu_{m-n}
d^\nu_n : +\frac{5}{8} \delta_{m,0}\;,
\nonumber\\
&~& F^{(\alpha , d)}_m = \sum_{n\in Z}G_{\mu \nu}
\alpha^\mu_{-n} d^\nu_{m+n}\;.
\eea
The open string metric explicitly appears in these operators. Similar 
relations also hold for the left parts of the virasoro operators i.e., for
${\tilde L}^{({\tilde \alpha} , {\tilde d})}_m$ and
${\tilde F}^{({\tilde \alpha} , {\tilde d})}_m$, and for the 
NS$\otimes$NS sector.

%%%%%%%%%%%%%%%%%%%%%%%%%%%%%%%%%%%%%%%%%%%%%%%%%%%%%%%%%%%%%%%%%%%%%%%%%%%%
{\bf $O(10,10;{\bf R})$ Duality relation}

Now we discuss $O(10,10;{\bf R})$ duality relation between the spacetime
and the closed string frame. In this subsection, let the metric
$g_{\mu\nu}$ be Euclidean. 
For $d$-dimensional toroidal compactification duality group is 
$ O(d,d; {\bf R})$ \cite{11,12}. The elements $h \in O(d,d; {\bf R})$
preserve the form of the matrix $J$ i.e.,
\bea
h^T J h =J \equiv 
 \left( \begin{array}{cc} 
 {\bf 0} & {\bf 1}_{d \times d} \\
 {\bf 1}_{d \times d} & {\bf 0}
\end{array} \right)\;.  
\eea

Define 20-dimensional vectors ${\bar R}$ and $R$ as
${\bar R}^m = \left( \begin{array}{c} 
{\bar X} \\
{\bar X'}
\end{array} \right)$ 
and
$ R^m = \left( \begin{array}{c} 
 X \\
 X'
\end{array} \right)$.  
The equations (10) and (11) imply that these vectors to have the 
relation ${\bar R}^m = \Gamma^{mn} R_n$, where 
$R_m = {\bar {\cal{G}}}_{mn}R^n$. The 
$20 \times 20$ matrices $\Gamma^{mn}$ and ${\bar {\cal{G}}}_{mn}$ are
\bea
 \Gamma^{mn} = \left( \begin{array}{cc} 
 G^{-1} & \frac{1}{2\pi \alpha'}\theta \\
 \frac{1}{2\pi \alpha'}\theta & G^{-1}
\end{array} \right)\;\;\;\;\;,\;\;\;\;\;  
{\bar {\cal{G}}}_{mn}= \left( \begin{array}{cc} 
 g & 0 \\
 0 & g
\end{array} \right)\;,
\eea
where $m,n \in \{1,2,...,20\}$.
The equations (10) and (11) also give ${\bar R}_m$ as 
${\bar R}_m = \Gamma_{mn} R^n$, where the matrix $\Gamma_{mn}$ is
\bea
\Gamma_{mn} = {\cal{G}}_{mp} \Gamma^{pq}{\bar {\cal{G}}}_{qn}
\;\;\;\;\;,\;\;\;\;\; 
{\cal{G}}_{mn}= \left( \begin{array}{cc} 
 G & 0 \\
 0 & G
\end{array} \right)\;.
\eea
The open string metric $G$ is, $G_{\mu\nu} = (g-(2\pi\alpha')^2
Bg^{-1}B)_{\mu\nu}$. Using the identity
$G^{-1}-\frac{1}{(2\pi \alpha')^2}\theta G \theta = g^{-1}$, we observe that,
the matrix $\Gamma$ is orthogonal and belongs to the duality group
$O(10,10;{\bf R})$. That is 
$(\Gamma^T)^{mp} \Gamma_{pn} = \delta^m_{\;\;\;\;n}$ and
$(\Gamma^T)^{mp} J_p^{\;\;\;q} \Gamma_{qn} = J^m_{\;\;\;\;n}$. 
In fact, these imply that the closed string frame and the spacetime are
related to each other by $O(10,10;{\bf R})$ duality rotation.

The square length of the vector ${\bar R}$ and the inner product
$G_{\mu\nu}{\bar X}^\mu{\bar X}'^\nu$ are preserved. Therefore, for the 
left and the right moving sectors of closed string, the square length also is
preserved i.e., $g_{\mu\nu}X^\mu_A X^\nu_A = G_{\mu\nu}{\bar X}^\mu_A 
{\bar X}^\nu_A$, where $A \in \{L,R\}$.

The equations (24) can be written as
$ \left( \begin{array}{c} 
 p \\
\frac{1}{\alpha'}L         
\end{array} \right)^m= 
\Gamma^{mn} \left( \begin{array}{c} 
 \Pi \\
\frac{1}{\alpha'}\Lambda
\end{array} \right)_n$. Note that $\frac{1}{\alpha'}L^\mu$ and
$\frac{1}{\alpha'}\Lambda^\mu$ are momenta of closed         
string in the dual spaces $\{{\bar X}'^\mu\}$ and $\{X'^\mu\}$, 
respectively. Therefore, the total momentum vectors in these 
$20$-dimensional spaces also
are related to each other by $ O(10,10; {\bf R})$ duality group.

The (10,10) Lorentzian momenta $(p^\mu_L , p^\mu_R)$, for $g_{\mu\nu}
= \delta_{\mu\nu}$ and $\alpha' = \frac{1}{2}$, form an even self-dual
Lorentzian lattice \cite {11}, i.e., the Lorentzian length is even
\bea
p^2_L - p^2_R = 2m^\mu n_\mu \;,
\eea
where $p^2_A = G_{\mu\nu}p^\mu_A p^\nu_A$ and $A\in \{L,R\}$. 
To obtain the equation (31), we assumed
that all directions of the spacetime to have the same radius of 
compactification. Note that all even self-dual $(d,d)$
Lorentzian lattices are related by $ O(d,d; {\bf R})$ rotations \cite{13}.
Since the momenta $(p^\mu_L , p^\mu_R)$ transform 
as vectors under $ O(10,10; {\bf R})$
group, Hamiltonian in the closed string frame is invariant.
More properties of the above duality group can be found in 
the Ref.\cite{14}.

%%%%%%%%%%%%%%%%%%%%%%%%%%%%%%%%%%%%%%%%%%%%%%%%%%%%%%%%%%%%%%%%%%%%%%%%%%%%
{\bf Quantization}

The quantization of the bosonic part gives the following equations
\bea
[ \xi^\mu , \Pi^\nu] = [ \xi'^\mu , \frac{1}{\alpha'} \Lambda^\nu ] 
= i g^{\mu \nu}\;,
\eea
which can be written as
\bea
[ x^\mu , p^\nu ] = [ x'^\mu , \frac{1}{\alpha'} L^\nu ] = i G^{\mu \nu}\;.
\eea
Also the oscillators satisfy the relations
\bea
[ \alpha^\mu_m , \alpha^\nu_n ] =
[ {\tilde \alpha}^\mu_m , {\tilde \alpha}^\nu_n ] = m \delta_{m+n,0} 
G^{\mu \nu}\;.
\eea

The quantization of the fermionic part leads to the relations 
\bea
&~& \{ b^\mu_r , b^\nu_s \} = \{ {\tilde b}^\mu_r , {\tilde b}^\nu_s \} = 
\delta_{r+s,0} G^{\mu \nu}\;,
\nonumber\\
&~& \{ d^\mu_m , d^\nu_n \} = \{ {\tilde d}^\mu_m , 
{\tilde d}^\nu_n \} = \delta_{m+n,0} G^{\mu \nu}\;.
\eea
where $(b^\mu_r \;,\; {\tilde b}^\mu_r)$ and $(d^\mu_n \;,\; 
{\tilde d}^\mu_n)$ are oscillators of the spinor ${\bar \psi}^\mu$ in 
the NS$\otimes$NS and R$\otimes$R sectors, respectively.
The quantization of the worldsheet fields in the closed string frame, 
explicitly depends on the open string metric.

Note that for the variables in the coordinate system $\{X^\mu\}$, raising,
lowering or contraction of indices can be done by the metrics 
$g_{\mu\nu}$ and $g^{\mu\nu}$, while for the variables in the 
closed string frame, such as $\{x^\mu, x'^\mu, p^\mu, L^\mu, \alpha^\mu_n,
b^\mu_r, d^\mu_m,...\}$, indices can be raised, lowered or contracted by
the open string metrics $G_{\mu\nu}$ and $G^{\mu\nu}$.
%%%%%%%%%%%%%%%%%%%%%%%%%%%%%%%%%%%%%%%%%%%%%%%%%%%%%%%%%%%%%%%%%%%%%%%%%%%%
\section{Boundary conditions of closed superstring}

Now we develop the boundary state formalism for a D$_p$-brane with 
background field. Assume that the $B$-field has non-vanishing components
only along the brane directions i.e., $B_{i \alpha}= B_{ij}=0$ and
$B_{\alpha\beta} \neq 0$.
Therefore, the bosonic boundary state equations are
\bea
&~&(G^{\alpha \beta}\partial_{\tau} X_\beta +\frac{1}{2\pi \alpha'}
\theta^{\alpha \beta}\partial_{\sigma}X_\beta)_{\tau=0} | B_b \rangle = 0\;,
\nonumber\\
&~&(\delta X^i)_{\tau = 0} | B_b \rangle = 0\;.
\eea
The set $\{X^\alpha\}$ shows the brane directions and the set 
$\{X^i\}$ shows the directions perpendicular to the brane.
Since the ghost and the super-ghost parts of the 
boundary state are independent of the background field, we do not study them.

The boundary conditions on the fermionic degrees of freedom should be 
imposed on both R$\otimes$R and NS$\otimes$NS sectors.
Worldsheet supersymmetry requires the two sectors to 
satisfy the boundary conditions                   
\bea
&~& \bigg{(}G^{\alpha \beta}(\psi_{-\beta} -i\eta \psi_{+\beta}) 
-\frac{1}{2\pi\alpha'}\theta^{\alpha \beta}(\psi_{-\beta} 
+i\eta \psi_{+\beta})\bigg{)}_{\tau=0} | B_f ,\eta \rangle =0\;,
\nonumber\\
&~& (\psi^i_- + i\eta \psi^i_+)_{\tau =0} 
| B_f ,\eta \rangle =0\;, 
\eea
where $\eta = \pm 1$ used to make GSO projection easily. 

In terms of oscillators the bosonic part of the boundary state equations 
becomes
\bea
(G^{\alpha \beta} \Pi_\beta + \frac{1}{2\pi \alpha'^2}\theta^{\alpha \beta}
\Lambda_\beta ) | B_b \rangle =0\;,
\eea
\bea
\Lambda^i | B_b \rangle =0\;,
\eea
\bea
(\xi^i - y^i ) | B_b \rangle =0\;,
\eea
\bea
\bigg{(}(G^{-1}-\frac{1}{2\pi \alpha'} \theta)^{\alpha\beta}a_{n \beta}  
+(G^{-1}+\frac{1}{2\pi \alpha'} \theta)^{\alpha\beta}
{\tilde a}_{-n \beta} \bigg{)} | B_b \rangle =0\;, 
\eea
\bea
(a^i_n - {\tilde a}^i_{-n}) | B_b \rangle =0\;.
\eea
The set $\{y^i\}$ indicates the transverse coordinates of the brane. 
The equations (36) and (37) reveal the effects of the noncommutativity
of the brane on the closed superstring boundary state.
The equation (38) describes the relation between the 
momentum (the momentum numbers) of the closed string and its winding 
numbers \cite{10}.
This equation implies that the noncommutativity and compactness of 
spacetime are coupled to each other. The lack of one of them leads to 
$\Pi^\alpha = 0$, which means that the emitted closed string propagates 
perpendicular to the brane. The second equation of (24) and the equation 
(38) give
\bea
p^\alpha = 0 \;,
\eea
that is, in the closed string frame, the closed string is emitted 
perpendicular to 
the brane. Also the first equation of (24) and the equation (39) lead to
\bea
L^i =0\;,
\eea
which means in the closed string frame, the closed string can not wrap 
around the compact direction ${\bar X}^i$.

The boundary state equations (36) and (37) in the closed string 
frame have the forms
\bea
&~&(\partial_{\tau} {\bar X}^\alpha)_{\tau=0} | B_b \rangle = 0\;,
\nonumber\\
&~&(\partial_\sigma {\bar X}^i)_{\tau = 0} | B_b \rangle = 0\;,
\nonumber\\
&~& ({\bar \psi}^\alpha_- -i\eta {\bar \psi}^\alpha_+)
_{\tau=0} | B_f ,\eta \rangle =0\;,
\nonumber\\
&~& ({\bar \psi}^i_- + i\eta {\bar \psi}^i_+)_{\tau =0} 
| B_f ,\eta \rangle =0\;. 
\eea
Apparently these equations have the form of the boundary state equations 
of the brane without background field.
Since in mode expansions of these equations the oscillators depend on the
background field, these equations describe the boundary conditions of a
closed string emitted from a brane with background field. The equations 
(43) and (44) directly can be extracted from the first and the second
equations of (45).
%%%%%%%%%%%%%%%%%%%%%%%%%%%%%%%%%%%%%%%%%%%%%%%%%%%%%%%%%%%%%%%%%%%%%%%%%%%%
\section{Conclusions}

We found a compact coordinate system 
(i.e., $\{ {\bar X}^\mu \}$) such that each coordinate and its T-dual 
coordinate are related to the spacetime coordinates and the T-dual 
coordinates of the spacetime by the 
noncommutativity. Similar relations also hold for the fermions 
of the worldsheet $\{{\bar \psi}^\mu \}$.
A novel feature is to cause the closed string state to have momentum which
is a linear combination of its spacetime momentum and its winding numbers.
In this frame the quantization of the worldsheet bosons and fermions is 
much similar to the zero $B$-field case 
that is, the closed string metric should be changed by the 
open string metric. This change also takes place for the supercurrent and
the superconformal generators. We saw that the spacetime and the closed
string frame are related to each other by $ O(10,10; {\bf R})$ duality
rotation. 

We observed the effects of noncommutativity and
compactification in the boundary state equations of a closed superstring
emitted from a noncommutative wrapped brane. In the closed string frame, the
closed string propagates perpendicular to the brane.
In this coordinate system the boundary conditions of the closed 
superstring appear similarly to the case without $B$-field. 
%%%%%%%%%%%%%%%%%%%%%%%%%%%%%%%%%%%%%%%%%%%%%%%%%%%%%%%%%%%%%%%%%%%%%%%%%%%%

\end{document}